\documentclass[aps,pra,amsmath,twocolumn,amssymb,footinbib,superscriptaddress]{revtex4-1}

\usepackage{amssymb,amsfonts,amsmath}

\usepackage{color}

\usepackage[apple mac]{inputenc}
\usepackage[english]{babel}
\usepackage{times}
\usepackage{latexsym}
\usepackage{fancyhdr}
\usepackage{verbatim}
\usepackage{graphicx}
\usepackage{epsf}
\usepackage{subfigure}
\usepackage{bm}
\usepackage{epstopdf}

\usepackage[colorlinks=true , citecolor=blue,urlcolor=blue]{hyperref}

\definecolor{Nathanblue}{rgb}{0.,0.24,0.51}
\newcommand{\blue}{\color{Nathanblue}}

\def\be{\begin{equation}}
\def\ee{\end{equation}}

\def\bs#1{\boldsymbol{#1}}

\begin{document}

\title{{\blue Topological Hofstadter Insulators in a Two-Dimensional Quasicrystal}}

\author{Duc-Thanh Tran}
\email{ducttran@ulb.ac.be}
\affiliation{Center for Nonlinear Phenomena and Complex Systems, Universit\'e Libre de Bruxelles, CP 231, Campus Plaine, B-1050 Brussels, Belgium}

\author{Alexandre Dauphin}
\email{adauphin@ulb.ac.be}
\affiliation{Center for Nonlinear Phenomena and Complex Systems, Universit\'e Libre de Bruxelles, CP 231, Campus Plaine, B-1050 Brussels, Belgium}\affiliation{Departamento de F\'isica Te\'orica I, Universidad Complutense, 28040 Madrid, Spain}

\author{Nathan Goldman}
\email{ngoldman@ulb.ac.be}
\affiliation{Center for Nonlinear Phenomena and Complex Systems, Universit\'e Libre de Bruxelles, CP 231, Campus Plaine, B-1050 Brussels, Belgium}
\affiliation{Laboratoire Kastler Brossel, CNRS, UPMC, ENS, Coll\`ege de France, 11 place Marcelin Berthelot, 75005, Paris, France}

\author{Pierre Gaspard}
\email{gaspard@ulb.ac.be}
\affiliation{Center for Nonlinear Phenomena and Complex Systems, Universit\'e Libre de Bruxelles, CP 231, Campus Plaine, B-1050 Brussels, Belgium}

\begin{abstract}

We investigate the properties of a two-dimensional quasicrystal in the presence of a uniform magnetic field. In this configuration, the density of states (DOS) displays a Hofstadter butterfly-like structure when it is represented as a function of the magnetic flux per tile. We show that the low-DOS regions of the energy spectrum are associated with chiral edge states, in direct analogy with the Chern insulators realized with periodic lattices. We establish the topological nature of the edge states by computing the topological Chern number associated with the bulk of the quasicrystal. This topological characterization of the non-periodic lattice is achieved through a local (real-space) topological marker. This work opens a route for the exploration of topological insulating materials in a wide range of non-periodic lattice systems, including photonic crystals and cold atoms in optical lattices.

\end{abstract}

\date{\today}

\maketitle

\section{Introduction}\label{sect:introduction}

Topological states of matter are classified according to topological invariants, which are intrinsic properties of  Bloch energy bands \cite{Hasan2010,Qi2011}.  Two-dimensional lattices subjected to a magnetic field display non-trivial topological bands characterized by the Chern or TKNN number, whose value is directly related to the quantized Hall conductivity \cite{Thouless1982,Kohmoto:1985}. Bloch bands of time-reversal-invariant topological insulators are associated with a $Z_2$ topological invariant \cite{Kane:2005bis}, which is rooted in the second Chern number \cite{Avron:1988,Qi2008,Fukui:2008}. While such topological indices can be related to the existence of edge-state transport, through the bulk-edge correspondence \cite{Hasan2010,Qi2011}, they constitute a property of the bulk. These numbers are defined in terms of integrals performed over the first Brillouin zone, which implicitly assumes that the system is translationally invariant, and more formally, that periodic boundary conditions have been applied. Generalizations of this approach have been proposed by Niu \emph{et al.} \cite{Niu:1985}, Bellissard \emph{et al.} and Prodan \cite{Bellissard:1994, Prodan:2010}, Bianco and Resta \cite{Bianco:2011gd} and Garcia \emph{et al.} \cite{Garcia:2014ts}, to analyze the robustness of topological order in the presence of disorder (see also Refs. \cite{Xu:2006,Kitaev_2006}). The fate of topological insulators subjected to translational-symmetry-breaking terms is by no means obvious, as recently revealed through the topological Anderson insulator \cite{Li:2009,Groth:2009}. 

The search for topological properties in non-periodic lattice structures constitutes an active field of research. In this context, much interest has been set upon the possibility to generate topological states using low-dimensional quasicrystals. Indeed, it was shown that 1D quasicrystals could be mapped onto the dimensional reduction of the 2D Harper-Hofstadter model \cite{hof}, which exhibits non-zero Chern numbers and chiral edge states at its boundaries \cite{Kraus:2012prl,Madsen:2013,Deng:2014bk}. In particular, such structures have been realized using photonic quasicrystals \cite{Verbin:2013}. A generalization to explore $Z_2$ topological insulators with 1D quasicrystals has been proposed by Mei et al. \cite{Mei:2012}. Recently, Kraus et al. \cite{Kraus:2013} showed that 2D quasicrystals could be mapped unto a dimensional reduction of a 4D quantum Hall model \cite{Qi2008}.

In this article, we follow another approach and investigate the topological properties of a 2D quasicrystal subjected to a uniform magnetic field. We base our study on the generalized Rauzy tiling (GRT) \cite{Vidal:2001hr}, presented in Fig. \ref{figure:fig1}. Contrary to previous works \cite{Kraus:2012prl,Madsen:2013,Deng:2014bk,Mei:2012,Kraus:2013}, we do not relate this quasicrystal to the dimensional reduction of a higher-dimensional lattice model. Instead, we show that the 2D quasicrystal shares the topological properties of a 2D Chern insulator, namely, it is characterized by non-zero Chern numbers and chiral edge states at its 1D boundary. Our results provide a novel interpretation for the Hofstadter butterfly-like spectrum associated with the quasicrystal \cite{vidal2004}: by analyzing the topological order of the system, we show that this fractal energy spectrum hosts a variety of Chern (Hofstadter) insulators \cite{Osadchy:2001wc}. Contrary to perturbative approaches aiming to study the robustness of quantum Hall phases in the presence of weak disorder, the Chern insulating phases revealed in this work are associated with a genuine non-periodic 2D system (i.e.~which cannot be smoothly connected to a periodic one). \\

The rest of the paper is organized as follows. Section~\ref{sect:The model} describes the construction of the quasicrystal; it also defines the tight-binding description in the presence of a uniform magnetic field, including the Peierls phase-factors~\cite{hof}. The Section~\ref{sect:The energy spectrum} discusses the energy (Hofstadter-like) spectrum and density of states of the quasicrystal for different boundary conditions.  The existence of chiral edge states is demonstrated in Section~\ref{section:edge}. The topologically invariant Chern number is evaluated in Section~\ref{sect:Computation of the first Chern number}, where a special emphasis is set upon the real-space Chern-number calculation introduced by Bianco and Resta \cite{Bianco:2011gd}. The last Section~\ref{section:conclusion} is dedicated to concluding remarks.

\section{The model}\label{sect:The model}

\begin{figure}
\begin{center}
\includegraphics[width=5cm]{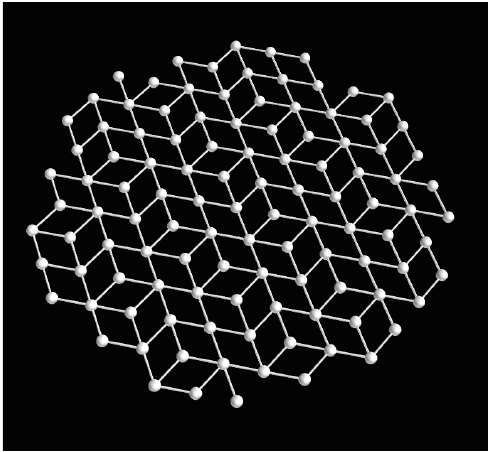}
\end{center}
\caption{A piece of the 2D isometric generalized Rauzy tiling (iGRT). }
\label{figure:fig1}
\end{figure}

We start with the two-dimensional generalized Rauzy tiling (GRT)~\cite{Vidal:2001hr}, presented in Fig.~\ref{figure:fig1}. The latter is constructed using the so-called cut-and-project method~\cite{Vidal:2001hr,vidal2004}: as explained below, the standard three-dimensional cubic lattice is projected into a chosen plane, following a selection rule based on the generalized Fibonacci sequence. Importantly, the cut-and-project method generates a finite-size tiling, called the approximant of order $\mathcal{O}$, whose infinite expansion generates the full quasi-periodic tiling. For the sake of completeness, we now sketch the cut-and-project method in the following Section \ref{app:qc}, see also Refs. \cite{Vidal:2001hr,vidal2004}. This Section also introduces the isometric generalized Rauzy tiling (iGRT) \cite{vidal2004},  where all the nearest-neighboring links share the same length. In this setting, all the tiles of the quasi-periodic tiling also have the same area [Fig.~\ref{figure:fig1}], which is particularly suitable for the study of quasicrystals subjected to magnetic fields. The magnetic field is introduced in Section ~\ref{app:peierls}, where we discuss the properties of Peierls phase-factors~\cite{hof} in a tight-binding description of the quasicrystal.

\subsection{\label{app:qc} Construction of the isometric generalized Rauzy tiling}

We start with a 3D cubic lattice of spacing $a$, and we generally choose $a=1$ as our unit length. The construction of the 2D generalized Rauzy quasicrystal is based on a ``projection" plane $\Pi_\theta$, which is perpendicular to the vector $\mathbf{B}_{\perp,\theta}=(1,\theta^{-1},\theta^{-2})$, and which contains the origin $(0,0,0)$. The number $\theta$ is chosen to be the \emph{Pisot root}, i.e., the real solution to the cubic equation $x^3=x^2+x+1$. As will be discussed below, the vector $\mathbf{B}_{\perp,\theta}$ is intimately related to the Fibonacci sequence. The generalized Rauzy quasicrystal is  constructed according to the following rules: 
\begin{enumerate}
\item{} The unit cube of the initial cubic lattice, of volume $a^3$, is slid along the plane $\Pi_\theta$. The resulting volume is referred to as the slice $\mathcal{S}$.
\item{}  All the lattice sites within the slice $\mathcal{S}$ are then projected orthogonally unto the projection plane $\Pi_\theta$.
\item{} The nearest-neighbor links of the quasicrystal are finally obtained through the projection of the nearest-neighbor links defined within the slice $\mathcal{S}$.
\end{enumerate}
This construction generates an infinite quasicrystal. In order to obtain a finite sample, an \emph{approximant} of the quasicrystal can be obtained through the cut-and-project method \cite{Vidal:2001hr}. This approximate scheme is based on the generalized Fibonacci sequence
\begin{equation}
F_n=F_{n-1}+F_{n-2}+F_{n-3}\text{,}
\end{equation}
where $F_{-1}=0$ and $F_0=F_1=1$. This sequence allows one to define a vector
\begin{equation}
\mathbf{B}_{\perp,n}=(F_n,F_{n-1},F_{n-2})\text{,}
\end{equation}
which converges towards a vector proportional to $\mathbf{B}_{\perp,\theta}$, defined above in terms of the Pisot root $\theta$, in the limit $n\rightarrow \infty$. Hence, it is natural to introduce the corresponding projection plan $\Pi_n$, which is perpendicular to $\mathbf{B}_{\perp,n}$ and contains the origin $(0,0,0)$. For a given integer $n$, the projection plane $\Pi_n$ allows one to define a finite tilling,  called the approximant of order $\mathcal O =n$, and which is constructed based on the rules given above. To be explicit, the tiling lies within the projection plane $\Pi_n$, and it is delimited by the parallelogram defined by the in-plane vectors 
\begin{align}
&\mathbf{B}_{1,n}=(F_{-n-1},F_{-n-2}+F_{-n-3},F_{-n-2}) ,\\
&\mathbf{B}_{2,n}=(F_{-n-2},F_{-n-3}+F_{-n-4},F_{-n-3})\text{.}
\end{align}
The slice $\mathcal S_n$, introduced above for $n \rightarrow \infty$, is obtained by sliding the unit cube along the parallelogram generated by $\mathbf{B}_{1,n}$ and $\mathbf{B}_{2,n}$. Finally, the lattice sites located within the slice $\mathcal S_n$ are projected unto the plane $\Pi_n$. \\

\begin{figure}
\begin{center}
\includegraphics{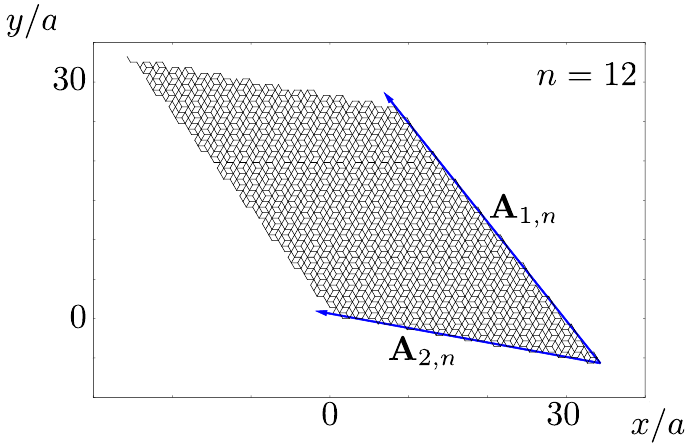}
\end{center}
\caption{Approximant for the isometric generalized Rauzy tilling (iGRT) of order $\mathcal O=n=12$. The approximant is delimited by the parallelogram generated by the vectors $\mathbf{A}_{1,n}$ and $\mathbf{A}_{2,n}$, see text.}
\label{figure:app_a1_a2}
\end{figure}

In order to simplify the following study,  it is convenient to build a quasicrystal displaying tiles of equal area. This can be realized by introducing an additional projection plane, which we choose to be perpendicular to the vector $(1,1,1)$. Projecting the vectors $\mathbf{B}_{1,n}$ and $\mathbf{B}_{2,n}$ unto this plane, yields the new vectors $\mathbf{A}_{1,n}$ and $\mathbf{A}_{2,n}$. The latter delimits a parallelogram in the new projection plane, within which the  (projected) approximant of the quasicrystal finally lies, see Fig. \ref{figure:app_a1_a2}. In this final configuration, all the nearest-neighbor links share the same length, $l=\sqrt{2/3}a $, and the quasicrystal is covered by tiles of area $A_\text{tile}=l^2 \sqrt{3}/2$. The resulting quasicrystal, illustrated in Figs. \ref{figure:fig1} and \ref{figure:app_a1_a2}, is called the \emph{isometric generalized Rauzy tiling} (iGRT).

\subsection{\label{app:peierls} Tight-binding Hamiltonian \\ and Peierls phase-factors in the quasicrystal}

%

We now introduce the single-particle Hamiltonian describing the iGRT in the presence of a uniform magnetic field applied perpendicularly to the quasicrystal.  In a tight-binding description, the hopping matrix elements between nearest neighbors acquire Peierls phase-factors~\cite{hof}, so that the Hamiltonian for spinless fermions is taken in the form
\begin{align} 
\hat{H}= - J \sum_{\langle j,k \rangle}e^{i \theta_{jk}} \hat c^{\dagger}_k \hat c_j, \quad \theta_{jk}= \int_{\bs{r}_j}^{\bs{r}_k} \mathbf{A}\cdot d\mathbf{l} ,
\label{eq1}
\end{align}
where $c^\dagger_j$ creates a fermion at the lattice site $\bs r_j$, $J$ is the hopping matrix element, $\exp (i \theta_{jk})$
denotes the Peierls phase-factor due to the magnetic field \cite{hof}, and $\mathbf{A}$ is the corresponding vector potential. The magnetic flux per tile is $\phi=B l^2 \sqrt{3}/2$ over the entire quasicrystal, where $l$ is the distance between neighboring sites. The flux quantum equals $\phi_0=2 \pi$ in the present units where $\hbar\!=\!e\!=\!1$.

\begin{figure}
\begin{center}
\includegraphics{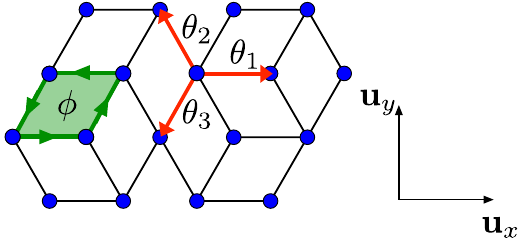}
\end{center}
\caption{The three possible directions for the links between nearest neighbors, and their corresponding Peierls phase-factors. Note that the sites of the quasicrystal can potentially have up to 5 nearest neighbors. The ``unit" vectors $\mathbf{u}_x$ and $\mathbf{u}_y$ have length $a= \sqrt{3/2} l$, where $l$ is the unit length of the quasicrystal, and where $a$ is the unit length on the initial cubic lattice. The unit vector $\mathbf{u}_x$ is chosen along one of the links, hereafter denoted as the $x$ direction.}
\label{figure:app_peierls}
\end{figure}

The Peierls phase-factors in Eq.~\eqref{eq1} depend on the direction of the links. Considering a lattice site at $\mathbf{r}_i$, there are potentially six types of Peierls phase-factors, with arguments $\pm\theta_1(\mathbf{r}_i)$, $\pm\theta_2(\mathbf{r}_i)$ and $\pm\theta_3(\mathbf{r}_i)$, see Fig. \ref{figure:app_peierls}. Choosing the Landau gauge, their expressions are explicitly given by
\begin{align}
\theta_1(\mathbf{r}_i)=0 , \, \, \theta_2(\mathbf{r}_i)=\phi \left ( \dfrac{x_i-l/4}{l} \right )= -\theta_3(\mathbf{r}_i),
\end{align}
where $l$ is the unit length of the quasicrystal, and $\phi$ is the flux penetrating each tile.

%

\section{The spectrum and density of states}\label{sect:The energy spectrum}

In the case of a 2D square lattice, the presence of a uniform magnetic flux $\phi$ opens  gaps in the energy spectrum. This gives rise to the celebrated Hofstadter's butterfly~\cite{hof}, which exhibits a fractal structure in the energy-flux plane. The Hofstadter bands are usually associated with non-zero Chern numbers, hence leading to Chern insulating phases \cite{Thouless1982}. Accordingly,  considering open boundary conditions, edge states at energies   located within the bulk energy gaps propagate along the boundary in a chiral manner \cite{Hatsugai:1993}. In this article, we investigate how such topological properties arise in the context of a 2D quasicrystal. 

\subsection{Open boundary conditions}

\begin{figure}
\begin{center}
\includegraphics[width=8cm]{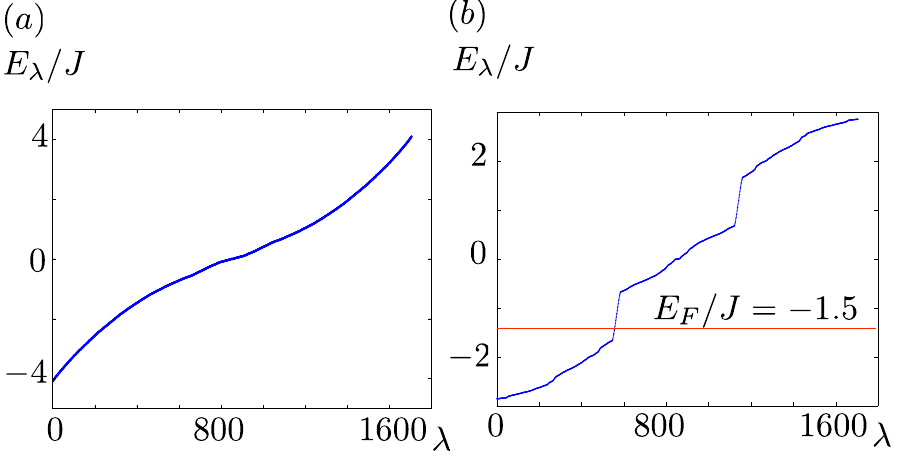}
\end{center}
\caption{(a) The energy spectrum $E_{\lambda}$ for the approximant of order $\mathcal O=12$ (1705 sites) in the absence of magnetic field versus the eigenvalue number $\lambda$.  (b) Spectrum for the same approximant subjected to a magnetic flux per unit tile $\phi/\phi_0=1/3$. The energy unit is the tight-binding tunneling matrix element $J$.}
\label{figure:figenergy}
\end{figure}

In the limit $\phi \rightarrow 0$, the Rauzy tiling in Fig.~\ref{figure:fig1} displays a gapless energy structure, shown in Fig.~\ref{figure:figenergy}~(a), and which leads to the unusual conductivity properties reported in Ref. \cite{Triozon:2002ja}. When adding a constant magnetic field, the density of states (DOS) is strongly reduced within large intervals. Vidal and Mosseri showed that these regions of reduced DOS also form a butterfly-like pattern in the $E$-$\phi$ plane \cite{vidal2004}. In the present work, we investigate the properties of these low-DOS regions in detail, and reveal their topological nature.

\begin{figure}[!]
\begin{center}
\includegraphics[width=8.75cm]{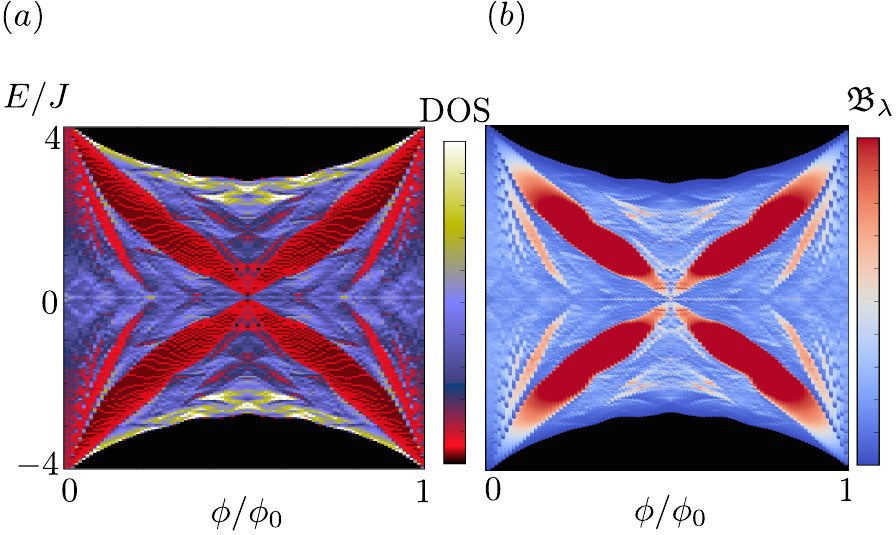}
\end{center}
\caption{(a) Density of states (DOS) in terms of the magnetic flux $\phi$ for the approximant of order $\mathcal O=12$. The low-DOS regions of the spectrum correspond to the spectral gaps in the closed (torus) geometry. (b) The edge-locality marker $\mathfrak{B}_{\lambda}=\sum_{\mathbf{r} \in \text{edge}}|\psi_{\lambda}(\mathbf{r})|^2$ as a function of the flux $\phi$ and eigenenergy $E_{\lambda}$, where $\psi_{\lambda}$ is the eigenvector associated with the energy $E_{\lambda}$. The states lying in the low-DOS regions of (a) are located at the boundary of the 12th-order approximant.}
\label{fig:butterfly_density}
\end{figure}

We study the approximant of order $\mathcal{O}=12$ (1705 sites) with open boundary conditions (OBC). Figure~ \ref{fig:butterfly_density} (a) depicts the DOS in terms of the magnetic flux $\phi$. This figure has regions of very low density (in red),  in agreement with the result of Ref. \cite{vidal2004}. Moreover, it shows similarities with the Hofstadter's butterfly~\cite{hof} in the $E$-$\phi$ plane: the figure is symmetric with respect to the axes $E=0$ and $\phi=1/2$, and the main ``gaps" survive for wide ranges of the flux. 

\subsection{Periodic boundary conditions}\label{section:periodic}

It is instructive to compare these results with the ones obtained using periodic boundary conditions (PBC), i.e. by setting the quasicrystal onto a torus. This closing procedure is possible by interpreting the approximant as a ``magnetic supercell"  \cite{Thouless1982,Kohmoto:1985}. For a fixed order $\mathcal{O}$, this closing can only be realized for specific values of the flux, which is due to the space dependence of the Peierls phase-factors in Eq. \eqref{eq1}. The ``magnetic supercell" can be defined over the quasicrystal according to the periodicity conditions
\begin{equation}
\theta_{k}(\mathbf{r}_i+\mathbf{A}_{j,n})=\theta_k(\mathbf{r}_i) \text{ mod } 2 \pi, \, \text{ where }  k=1,2,3 \, \text{; } j=1,2 . \label{per_cond_theta}
\end{equation}
Here, the vectors $\mathbf{A}_{j,n}$ delimit the size of the quasicrystal, as described in Section \ref{app:qc}, see  also Fig.~\ref{figure:app_a1_a2}. Using the fact that the Peierls phase-factors only depend on the $x$-coordinate, together with Eq. \eqref{per_cond_theta} and the fact that 
\begin{align}
\mathbf{A}_{1,n}\cdot\mathbf{u}_x=\dfrac{l}{2} \times M , \, \, \mathbf{A}_{2,n}\cdot\mathbf{u}_x=\dfrac{l}{2} \times N \text{,} \quad M,N \in \mathbb{Z}, \label{eq:m_n}
\end{align}
we find that the magnetic flux should satisfy the following condition 
\begin{align}
\phi/\phi_0=\dfrac{2}{\text{gcd}(|M|,|N|)}\text{,} \label{cond_per}
\end{align}
for the periodic boundary conditions to be applied; here the notation ``gcd" stands for greatest common divisor. For the approximant of order $\mathcal O =12$, Eq. \eqref{eq:m_n} leads to the integers $M=-85$ and $N=35$, such that Eq. \eqref{cond_per} implies the specific value $\phi=2/5$ for the magnetic flux. In other words, the bulk properties of the quasicrystal subjected to a magnetic flux $\phi=2/5$ can be obtained by applying periodic boundary conditions to the approximant of order $\mathcal O =12$. For the approximant of order $\mathcal O =5$, Eq. \eqref{eq:m_n} yields $M=9$ and $N=-12$, so that the periodicity condition  \eqref{cond_per} fixes the flux to the value $\phi=2/3$. 

Using these values of the flux $\phi$ [i.e. $\phi/\phi_0=2/5$ and $\phi/\phi_0=2/3$], we have verified that the low-DOS regions of the spectrum obtained in the open geometry (OBC) correspond to \emph{energy gaps} in the PBC configuration. This analysis validates the intuitive fact that there is a one-to-one correspondence between the low-DOS regions of the open system and the bulk energy gaps captured by the torus geometry. We stress that the size of the main bulk gaps are of the order of $J$, indicating that the non-periodicity of the quasicrystal does not significantly reduce the gaps with respect to the regular square-lattice case \cite{hof}.

\section{Chiral edge states in the quasicrystal}\label{section:edge}

In analogy with the standard square-lattice case \cite{Hatsugai:1993}, we expect the low-DOS regions of the butterfly in Fig. \ref{fig:butterfly_density}~(a) to host (chiral) edge states \cite{vidal2004}. In order to validate this hypothesis, we present in Fig. \ref{fig:butterfly_density}~(b) the edge-locality marker
\begin{align}
\mathfrak{B}_{\lambda}=\sum_{\mathbf{r} \in \text{edge}}|\psi_{\lambda}(\mathbf{r})|^2 ,
\end{align}
as a function of the energy $E_{\lambda}$ and flux $\phi$. The quantity $\mathfrak{B}_{\lambda}$ characterizes the localization of each energy eigenstate $\psi_{\lambda}$, with energy $E_{\lambda}$, at the boundary of the approximant. We find that the corresponding plot  [Fig. \ref{fig:butterfly_density} (b)] follows the shape of the butterfly in Fig. \ref{fig:butterfly_density} (a). Indeed, each low-DOS region is shown to host edge states, which is in agreement with the fact that these regions correspond to bulk energy gaps when applying periodic boundary conditions (see above).

\begin{figure}
\begin{center}
\includegraphics[width=8.5cm]{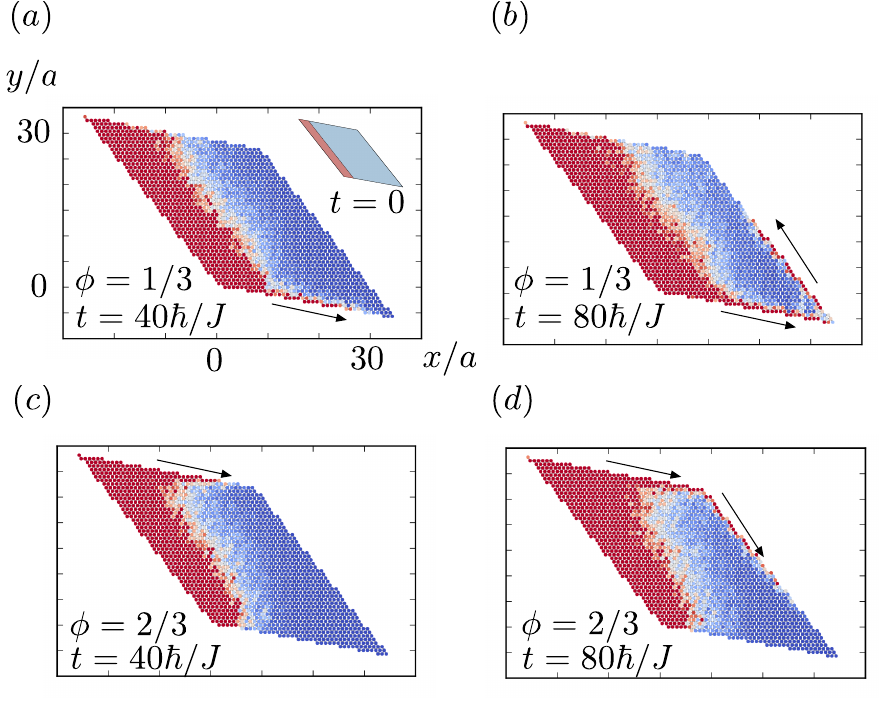}
\end{center}
\caption{Edge-state propagation for opposite magnetic flux: (a) A Fermi gas is initially confined for time $t<0$ at the left border of the quasi-crystal, using an infinite potential $V_\text{conf}=\infty$ [see the small inset in (a)]. The system has 1705 sites and we set the Fermi energy $E_F=-0.8J$ to be within the first energy gap (i.e. in the low-DOS region of the spectrum).  At time $t=0$, the confining potential is decreased to a lower value $V_\text{conf}=V_1=0.5J$, so as to allow edge-state motion while limiting the diffusion of the bulk states. Shown is the particle density $\rho (\bs x)$, with a colormap displaying populated regions in red. The simulation is performed for two ``opposite" values of the magnetic flux: (a)-(b) $\phi/\phi_0=1/3$ and (c)-(d) $\phi/\phi_0=2/3\equiv -1/3$, leading to opposite edge-state chiralities. Here, the unit length $a$ is defined as $a=\sqrt{3/2} l$, where $l$ is the nearest-neighbor distance in the quasicrystal. }
\label{fig:edge_evolution}
\end{figure}

We now analyze the chiral property of these gapless edge states, in view of offering a first signature for the existence of Chern insulating phases in 2D quasicrystals. Different methods could be exploited to highlight the presence of chiral edge states, such as transport measurement \cite{vonKlitzing:1986}, spectroscopy \cite{Liu:2010,Stanescu2010,Goldman:2012prl}, Aharonov-Bohm interferometry \cite{Ji:2003ck}, or \emph{in situ} imaging \cite{goldman_pnas}.  Here, for the sake of illustration, we consider the edge-filter method introduced in Ref. \cite{goldman_pnas}, which allows for a direct \emph{in situ} visualization of the edge-state propagation along the boundaries. For this purpose, we initially prepare the Fermi gas in a small region located at the left border of the quasicrystal, using a sharp potential $V_\text{conf}=\infty$, and we set the Fermi energy within the first bulk gap. At time $t>0$, we partially release the confining walls $V_\text{conf}=V_1$, and study the dynamics of the cloud, which is then allowed to propagate over the entire approximant.  The value $V_1 \approx W$ is chosen to be close enough to the lowest band's width $W$, so as to limit the diffusion of the bulk states, hence highlighting the edge-state motion. Indeed, in this configuration, bulk states are essentially trapped, while the edge states  living in the spectral bulk gap freely propagate. Figure \ref{fig:edge_evolution} presents the corresponding simulation, for a quasicrystal of order $\mathcal{O}=12$. Figures (a)-(b) plot the particle density at successive times, $t=40\hbar/J$ and $t=80\hbar/J$, for $\phi=1/3$. It unambiguously shows the counter-clockwise propagation of the edge states along the boundary. Figures (c)-(d) present similar dynamics for another value of the flux $\phi=2/3$, i.e. the time-reversal counterpart of $\phi =1/3$. In this latter case, the dynamics show a clockwise propagation of the edge states. These results demonstrate that the spectral gaps of the quasicrystal indeed host chiral edge states, which suggests that this non-periodic physical system could be associated with a non-trivial topological order (i.e. a non-zero Chern number \cite{Thouless1982,Kohmoto:1985}).

\section{Computation of the Chern number}\label{sect:Computation of the first Chern number}


We now fully characterize the topological nature of the system. We consider that the Fermi energy $E_{\text F}$ is set within the $r$th  gap of the bulk spectrum discussed above, in which case the topological order of the system is given by the ``total" Chern number $\nu_{\text{tot}} =\sum_{\alpha=1}^r \nu^{\alpha}_{\text{Ch}}$, where $\nu^{\alpha}_{\text{Ch}}$ are the Chern numbers of the individual (occupied) bands $\alpha$.  A first method consists in studying the system with periodic boundary conditions [Section~\ref{section:periodic}], in which case the  Chern numbers $\nu^{\alpha}_{\text{Ch}}$ can be computed using standard numerical methods developed in the context of Chern insulators. Applying the numerical algorithm of Fukui \emph{et al.} \cite{Fukui:2005ua,Hatsugai:2006wj}, and setting the magnetic flux to the values $\phi/\phi_0=2/5$  and $\phi/\phi_0=2/3$, we find that the two main wings of the butterfly-like structure in Fig. \ref{fig:butterfly_density} are associated with opposite  Chern numbers $\nu_{\text{tot}}=\pm 1$, respectively. In this regime, the 2D quasicrystal with periodic boundary conditions is hence topologically equivalent to the Hofstadter model, which is based on the (space-periodic) square lattice \cite{Thouless1982,Kohmoto:1985}. In particular, the lowest bulk band of the spectrum is associated with a nonzero Chern number $\nu^{1}_{\text{Ch}}=-1$ (resp. $\nu^{1}_{\text{Ch}}=+1$) when the flux is $\phi/\phi_0=1/3$ (resp. $\phi/\phi_0=2/3$).

\section{Real space characterization of the topological order}

The numerical estimation of the topological order described above, and which is based on the application of periodic boundary conditions, provides an indication that the chiral states living within the low-DOS regions of the spectrum are associated with a topological Chern number. However, this method relies on the restoration of spatial periodicity, which is implicit when imposing the closed boundary conditions. In particular, when considering a specific approximant for the quasicrystal, such periodic boundary conditions can only be applied  for given values of the magnetic flux (see Section~\ref{section:periodic}). Altogether, this suggests that it would be more satisfactory to evaluate the topological invariant of the system without invoking any (artificial)  restoration of spatial periodicity. This can be achieved using the \emph{real-space Chern number} introduced by Bianco and Resta in Ref. \cite{Bianco:2011gd}, which provides a local characterization of the bulk topological invariant, and hence, which is independent of the boundary conditions. 


\subsection{The method of Bianco-Resta}\label{ref_appendix_bianco}

In this Section~\ref{ref_appendix_bianco}, we  remind the general definition of the \emph{real-space Chern invariant} $\mathcal{C}$, as introduced by Bianco and Resta in Refs. \cite{Bianco:2011gd,Bianco_thesis}. This quantity can be exploited to characterize the topology of finite-size systems, locally in real-space, and it can thus be applied to any type of geometry and boundary conditions.  We first recall its properties in the case of translationally invariant systems, both for periodic and open boundary conditions. We then apply this method to quasicrystals in Section~\ref{ref_appendix_bianco_quasi}, and we provide the explicit expressions that are used in our numerical simulations. Numerical results are presented in Section~\ref{section:numerics}.


Let us first consider a portion of a translationally invariant crystal, i.e. a supercell of area $A$ constituted of several unit cells, on which we apply periodic boundary conditions. The corresponding eigenenergies and eigenstates are written as $E_{\lambda}$ and $\vert \psi_\lambda\rangle$, respectively. The Fermi energy $E_{\text F}$ will be assumed to lie within a band gap, in which case the topology of the system can be characterized by the ``total" Chern number
\be
\nu_{\text{tot}} = \sum_{\alpha}  \nu^\alpha_\text{Ch}, \label{def:total_chern}
\ee
where $\nu^\alpha_\text{Ch}$ denotes the Chern number \cite{Thouless1982,Kohmoto:1985} of the filled band $\alpha$ (with energies $E_{\lambda}  <E_{\text F} \in \alpha$).

We  introduce the Chern marker operator \cite{Bianco:2011gd,Bianco_thesis}
\begin{equation}
\hat{\mathfrak{C}}= -4\pi \text{Im} \left [ \hat{x}_{\mathcal{Q}} \,  \hat{y}_{\mathcal{P}} \right ]\text{,} \label{def:marker}
\end{equation}
where the operators $\hat{\mathbf{r}}_{\mathcal{P}}\!=\!\hat{\mathcal{P}} \, \hat{\mathbf{r}} \, \hat{\mathcal{Q}} \!=\! (\hat x_{\mathcal{P}},\hat y_{\mathcal{P}})$ and $\hat{\mathbf{r}}_{\mathcal{Q}}\!=\!\hat{\mathcal{Q}} \,\hat{\mathbf{r}} \, \hat{\mathcal{P}}\!=\! (\hat x_{\mathcal{Q}},\hat y_{\mathcal{Q}})$ are expressed in terms of the position operator $\hat{\mathbf{r}}\!=\!(\hat x,\hat y)$  and  the projection operators
\begin{equation}
\hat{\mathcal P}=\sum_{E_\lambda<E_{\text F}} \vert \psi_\lambda \rangle \langle \psi_\lambda \vert = \hat 1 - \hat{\mathcal Q} .\label{def:projectors}
\end{equation}
The operator $\hat{\mathcal P}$ projects onto the ground state of the Hamiltonian $\hat H$. The Chern marker operator $\hat{\mathfrak{C}}$ in Eq. \eqref{def:marker} is Hermitian and it commutes with the lattice-translation operators \cite{Bianco:2011gd,Bianco_thesis}. Besides, the operators $\hat{\mathbf{r}}_{\mathcal{P}, \mathcal{Q}}$ also commute with the lattice-translation operators, and they satisfy the locality property \cite{Bianco_thesis}
\be
\langle \mathbf{r} \vert \hat{\mathbf{r}}_{\mathcal{P}, \mathcal{Q}} \vert \mathbf{r}' \rangle \sim \exp ( - \kappa_{\mathcal{P}, \mathcal{Q}} \Vert \mathbf{r} -\mathbf{r}' \Vert), \label{locality}
\ee
where $\vert \mathbf{r} \rangle$ are the eigenstates of the usual position operator $\hat{\mathbf{r}}$, and where $\kappa_{\mathcal{P}, \mathcal{Q}} >0$.


The topology of the system can be related to the trace of the Chern marker operator $\hat{\mathfrak{C}}$ in Eq. \eqref{def:marker}. This can be shown by performing this trace using the Bloch states basis, and dividing by the supercell area $A$, which yields \cite{Bianco_thesis}
\begin{align}
\langle \hat{\mathfrak{C}} \rangle&= \frac{1}{A} \text{Tr} \,  \hat{\mathfrak{C}} \label{trace_bloch}\\
&=-\frac{2\pi i}{A} \sum_{\alpha} \sum_{\mathbf{k}} \left [\langle \partial_{k_x}\psi_{\alpha\mathbf{k}} \vert \partial_{k_y}\psi_{\alpha\mathbf{k}} \rangle- (k_x \leftrightarrow k_y) \right ] \text{,} \notag
\end{align} 
where $\psi_{\alpha\mathbf{k}}$ is the Bloch state with band index $\alpha$ and quasi-momentum $\mathbf{k}$. Importantly, the normalized trace in Eq. \eqref{trace_bloch} is the discretized form of a Chern number \cite{Thouless1982,Kohmoto:1985}.  
 Indeed this quantity tends towards the ``total" Chern number $\nu_\text{tot}$ in Eq. \eqref{def:total_chern} when taking the usual thermodynamic limit (TL):
\begin{equation}
\langle \hat{\mathfrak{C}} \rangle \xrightarrow{A \rightarrow \infty} \nu_\text{tot}=\sum_{\alpha} \nu^\alpha_\text{Ch}\text{.} \label{conv_chern}
\end{equation}

Besides, the normalized trace in Eq. \eqref{trace_bloch} can be evaluated in the position basis $\{ \vert \mathbf{r} \rangle \}$.  The expression for the normalized trace \eqref{trace_bloch} then reads
\begin{equation}
\langle \hat{\mathfrak{C}} \rangle=\frac{1}{A}\int_{\text{supercell}} \mathfrak{C} (\mathbf{r}') d\mathbf{r}'=\frac{1}{A_\text{cell}}\int_{\text{cell}} \mathfrak{C} (\mathbf{r}')d\mathbf{r}'\text{,} \label{trace_position}
\end{equation} 
where we introduced the local Chern marker 
\be
\mathfrak{C}(\mathbf{r})=\langle \mathbf{r} \vert \hat{\mathfrak{C}} \vert \mathbf{r} \rangle . \label{def_chern_marker}
\ee
The second equality in Eq. \eqref{trace_position} expresses the fact that the mean quantity $\langle \hat{\mathfrak{C}} \rangle$ can be equally obtained by restricting the spatial average over a \emph{single unit cell} of area $A_\text{cell}$, which is due to the fact that the Chern marker operator is invariant under lattice translations. Combining the latter result, together with the fact that the averaged trace $\langle \hat{\mathfrak{C}} \rangle$ converges towards the total Chern number \eqref{def:total_chern} (see Eqs. \eqref{trace_bloch}-\eqref{conv_chern}) motivates  the introduction of a \emph{real-space Chern number}, defined as
\be
 \mathcal{C} =\frac{1}{A_\text{cell}}\int_{\text{cell}} \mathfrak{C} (\mathbf{r}')d\mathbf{r}' . \label{def:real_space_chern}
\ee
The real-space Chern number $ \mathcal{C} $ provides an approximate value for the total Chern number $\nu_\text{tot}$ in Eq. \eqref{def:total_chern}, which becomes exact in the thermodynamic limit, $ \mathcal{C} \xrightarrow{A \rightarrow \infty} \nu_\text{tot}$. Although implicit in Eq. \eqref{def:real_space_chern}, the quantity $\mathcal{C}$ is \emph{local} in space, with a resolution given by the unit cell. However, we stress that this quantity contains information about the global system through the projectors in Eq. \eqref{def:projectors}. Note that the numerical calculation of the real-space Chern number requires a single diagonalization of the lattice Hamiltonian $\hat H$, in contrast with other methods based on momentum-space integration \cite{Fukui:2005ua}.

We now discuss how this method applies to the case where open boundary conditions are set upon the supercell. In this case, one verifies that the normalized trace in Eq. \eqref{trace_position} is trivial,
\be
\langle \hat{\mathfrak{C}} \rangle=\frac{1}{A}\int_{\text{supercell}} \mathfrak{C} (\mathbf{r}') d\mathbf{r}'=0, \label{trivial_av}
\ee
which is in agreement with the fact that the topology of a fiber bundle based over a flat manifold is necessarily trivial \cite{nakahara}. However, it is relevant to split the spatial average in Eq. \eqref{trivial_av} into two contributions,
\begin{align}
\langle \hat{\mathfrak{C}} \rangle=0&=\frac{1}{A}\int_{\text{bulk}} \mathfrak{C} (\mathbf{r}') d\mathbf{r}' + \frac{1}{A}\int_{\text{edge}} \mathfrak{C} (\mathbf{r}') d\mathbf{r}' \notag \\
&\equiv \langle \hat{\mathfrak{C}} \rangle_{\text{bulk}} + \langle \hat{\mathfrak{C}} \rangle_{\text{edge}}, \label{split_av}
\end{align}
which respectively correspond to contributions from the bulk and the edge of the supercell. Note that the edge width is considered to be finite. Invoking the locality property of the Chern marker operator $\hat{\mathfrak{C}}$, Eqs. \eqref{def:marker} and \eqref{locality}, it is reasonable to assume that the (local) real-space Chern number $\mathcal{C}$ in Eq. \eqref{def:real_space_chern} does not depend on the choice of boundary conditions if it is evaluated far away from the edges. Hence, the average over the bulk $\langle \hat{\mathfrak{C}} \rangle_{\text{bulk}} $ defined in Eq. \eqref{split_av} still provides an approximate value for the total Chern number. Invoking translational symmetry yields
\be
\langle \hat{\mathfrak{C}} \rangle_{\text{bulk}} = \frac{1}{A_\text{cell}}\int_{\text{cell}} \mathfrak{C} (\mathbf{r}')d\mathbf{r}' \xrightarrow{A \rightarrow \infty} \nu_\text{tot}, \label{local_bulk}
\ee
where the unit cell is supposed to be located far away from the edges. In particular, the latter result \eqref{local_bulk} shows that the topological order of the system can still be identified locally by the real-space Chern number $ \mathcal{C}$ in Eq. \eqref{def:real_space_chern}. In contrast to the case of periodic boundary conditions, the total average in Eq. \eqref{split_av} indicates that the edge is associated with an opposite contribution, which exactly compensates the topological order contained in the bulk. 

\subsection{Application to quasicrystals}\label{ref_appendix_bianco_quasi}

Having established the general theoretical framework, we now turn to the case of non-periodic systems. First of all, we note that the real-space quantity in Eq. \eqref{def:real_space_chern} is a well-defined quantity in non-periodic systems, as it is simply given by a spatial average of the local Chern marker \eqref{def_chern_marker} over a unit cell. However, the relation between the quantity in Eq. \eqref{def:real_space_chern} and the topological order of the system is no longer obvious in a non-periodic system, where the simplification presented in Eq. \eqref{trace_position} no longer holds. To address this question, it is convenient to start with a situation where the topological Chern number in Eq. \eqref{def:total_chern} is well defined. Let us first illustrate this approach with the typical example of a periodic lattice system subjected to a disordered potential \cite{Niu:1985,Bianco:2011gd}. Suppose that the Fermi energy is set within a bulk gap of the unperturbed lattice and that the disorder is progressively turned on. In this case, the bulk-averaged quantity $\langle \hat{\mathfrak{C}} \rangle_{\text{bulk}} $ defined in Eq. \eqref{split_av} will remain stable as long as the disorder is weak compared to the gap. This is due to the topological property of the Chern number $\nu_\text{tot}$ \cite{Niu:1985}, and the fact that   $\langle \hat{\mathfrak{C}} \rangle_{\text{bulk}} \xrightarrow{A \rightarrow \infty} \nu_\text{tot}$. When implementing the present method numerically for the case of disordered lattices, we find that the real-space Chern number in Eq. \eqref{def:real_space_chern}  fluctuates in a manner which depends on the degree of disorder. These fluctuations can be reduced by introducing a smoothened real-space Chern number, obtained by replacing the unit-cell average by an average over a disk $D$ of radius $r_D$, centered around $\mathbf{r}_0$ and located within the bulk,
\begin{equation}
\mathcal{C}_D(\mathbf{r}_0)=\frac{1}{A_D} \int_D \mathfrak{C}(\mathbf{r}') d\mathbf{r}'\text{,} \label{smooth_chern}
\end{equation}
where $A_D$ is the disk area. By taking the radius $r_D$ to be large compared to the fluctuation length, but sufficiently small to remain within the bulk, allows one to identify the topological order of the disordered system accurately. Note that the quantity $\mathcal{C}_D(\mathbf{r}_0)$ in Eq. \eqref{smooth_chern} is still local, and thus, it can be exploited to probe (potentially different) topological orders associated with connected subsystems. 

A similar analysis can be performed in the case of quasicrystals, which constitute the focus of the present work. As discussed in Section~\ref{sect:The energy spectrum}, quasicrystals subjected to a uniform magnetic field display a non-trivial spectral structure, constituted of separated bulk bands and gapless edge-states living within the bulk gaps. The bulk energy spectrum was obtained by applying periodic boundary conditions to a piece of the quasicrystal (i.e. to an approximant of order $\mathcal{O}$). For a given order $\mathcal{O}$, we recall that this ``closing" procedure can only be  realized for specific values of the flux $\phi$, for which the Peierls phase-factors are indeed periodic over the approximant (see Section \ref{app:peierls}). Choosing a matching pair $\mathcal{O}$-$\phi$, the periodic boundary conditions are well defined, which allows one to introduce the total Chern number $\nu_\text{tot}$ of the quasicrystal, for a given Fermi energy, see Eq. \eqref{def:total_chern}. In this configuration, the Bianco-Resta method discussed above is directly applicable to evaluate the quasicrystal topological order. For a regular (homogeneous) quasicrystal, such as the isometric generalized Rauzy tiling (iGRT) considered in this work, one verifies that the real-space Chern number $\mathcal{C}$ in Eq. \eqref{def:real_space_chern} already provides a good estimation of the topological Chern number. Note that in the quasicrystal framework, the average over the unit cell in Eq. \eqref{def:real_space_chern} should be replaced by an average over a single tile of the quasicrystal, and we remind that all the tiles have the same area in the quasicrystal. The real-space Chern number  \eqref{def:real_space_chern} has been evaluated numerically for the iGRT, and we indeed find that the result only slightly fluctuates from site to site, see Fig. \ref{fig:local} in the main text  and Eq. \eqref{lattice_chern} below. In order to inhibit any effects due to small-scale fluctuations, it is convenient to evaluate the smoothened real-space Chern number $\mathcal{C}_D(\mathbf{r}_0)$ in Eq. \eqref{smooth_chern}. Note that the latter marker is particularly suitable to evaluate the topological order of more irregular quasicrystals, where fluctuations of the real-space Chern number  \eqref{def:real_space_chern} are potentially large (similarly to the case of strongly disordered systems).

We now provide the explicit expressions used in a tight-binding description of the quasicrystal. The local Chern marker in Eq. \eqref{def_chern_marker}, defined at the lattice site $\mathbf{r}_i$, is written as
\begin{equation}
\label{eq:app:discrlocalcm}
\mathfrak{C}(\mathbf{r}_i)=-4\pi \text{Im} \left [\sum_{\mathbf{r}_j} \langle \mathbf{r}_i \vert \hat{x}_\mathcal{Q} \vert \mathbf{r}_j \rangle \langle \mathbf{r}_j \vert \hat{y}_\mathcal{P} \vert \mathbf{r}_i \rangle \right ]\text{,}
\end{equation}
where $\{ \vert \mathbf{r}_i \rangle \}$ denotes the lattice-position basis, and where
\begin{align}
 &\langle \mathbf{r}_i \vert \hat{x}_\mathcal{Q} \vert \mathbf{r}_j \rangle=\sum_{\mathbf{r}_k}  \mathcal{Q}(\mathbf{r}_i,\mathbf{r}_k)x_k \mathcal{P}(\mathbf{r}_k,\mathbf{r}_j) , \\
  &\langle \mathbf{r}_j \vert \hat{y}_\mathcal{P} \vert \mathbf{r}_i \rangle=\sum_{\mathbf{r}_k}  \mathcal{P}(\mathbf{r}_j,\mathbf{r}_k)y_k \mathcal{Q}(\mathbf{r}_k,\mathbf{r}_i) , \\
&\mathcal{P}(\mathbf{r}_i,\mathbf{r}_j)=\sum_{E_\lambda<E_F} \langle \mathbf{r}_i \vert \psi_\lambda \rangle \langle \psi_\lambda \vert \mathbf{r}_j \rangle\text{,} \\
&\mathcal{Q}(\mathbf{r}_i,\mathbf{r}_j)=\sum_{E_\lambda > E_F} \langle \mathbf{r}_i \vert \psi_\lambda \rangle \langle \psi_\lambda \vert \mathbf{r}_j \rangle\text{.}
\end{align}
The real-space Chern number in Eq. \eqref{def:real_space_chern} is also defined at a specific lattice site $\mathbf{r}_i$, and it is given by the averaged marker $\mathfrak{C}(\mathbf{r}_i)$ over a single quasicrystal tile, which in the present case simply reads 
\begin{equation}
\mathcal{C} (\mathbf{r}_j) = \frac{\mathfrak{C}(\mathbf{r}_j)}{A_\text{tile}} , \quad A_\text{tile}=l^2 \sqrt{3}/2 , \label{lattice_chern}
\end{equation}
where $A_\text{tile}$ is the tile area of the iGRT. Finally, the smoothened real-space Chern number $\mathcal{C}_D(\mathbf{r}_0)$ in Eq. \eqref{smooth_chern} is given by 
\begin{equation}
\mathcal{C}_D(\mathbf{r}_0)=\frac{1}{N}\sum_{j \in D} \mathcal{C} (\mathbf{r}_j) \text{,} \label{lattice_smooth}
\end{equation}
where the average is performed over the $N$ points contained in the disk $D$, centered at the site $\mathbf{r}_0$, with radius $r_D$. In practice, we choose the value of the radius $r_D$ depending on the size of the system and the typical fluctuation length.

\begin{figure}[!]
\begin{center}
\includegraphics[width=8.75cm]{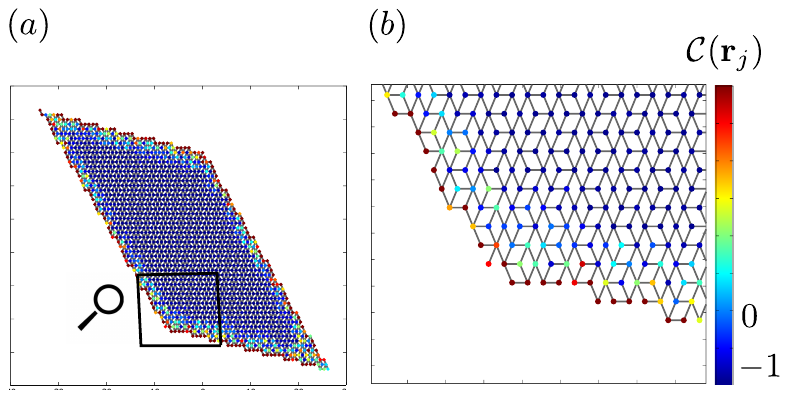}%
\end{center}
\caption{(a) Plot of the real-space Chern number $\mathcal{C} (\mathbf{r}_j)$ over the approximant of order $\mathcal{O}\!=\!12$. We set $E_{\text F}=-1.5 J$ within the first spectral bulk gap, as shown in Fig.~\ref{figure:figenergy} (b). The averaged Chern number $\mathcal{C}_D$ has also been evaluated within the bulk of the quasicrystal, considering a disk $D$ of radius $r_D \approx 6a$ (containing $186$ sites), yielding $\mathcal{C}_D \approx -1.002$. The fluctuation of the real-space Chern number $\mathcal{C} (\mathbf{r}_j)$ are found to be of the order of $10^{-2}$ within the bulk. A zoom of this figure is shown in (b).}
\label{fig:local}
\end{figure}

\begin{figure}[!]
\begin{center}
\includegraphics[width=8.5cm]{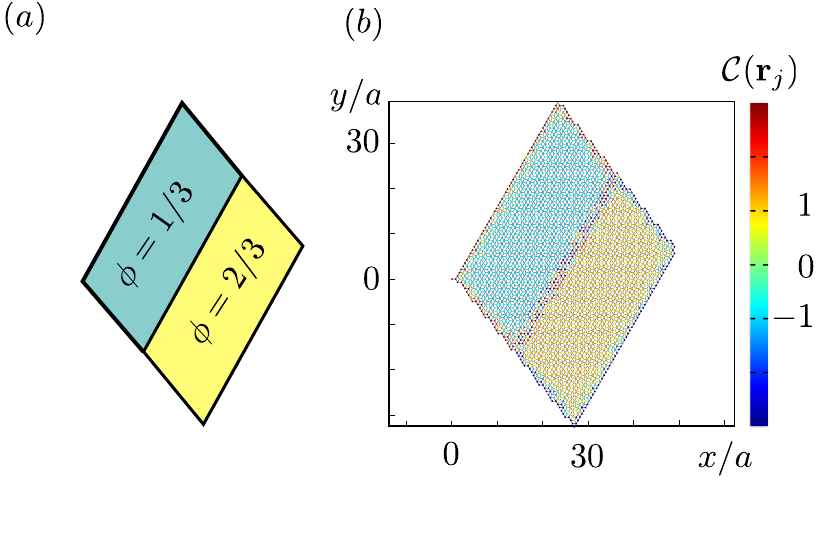}
\end{center}
\caption{(a) The heterojunction with fluxes $\phi/\phi_0=1/3$ and $\phi/\phi_0=2/3$ per tile. (b) Plot of the real-space Chern number $\mathcal{C} (\mathbf{r}_j)$, for a heterojunction based on the approximant of order $\mathcal{O}=13$. The Fermi energy $E_{\text{F}}=-1.5J$ is set within the first energy bulk gap. The averaged Chern number $\mathcal{C}_D$ has been evaluated within the two regions, considering two different disks $D$ of radius $r_D \!\approx\! 4 a$ (i.e. containing 98 sites each), centered within the two corresponding regions. This yields $\mathcal{C}_D(\mathbf{r}_1) \approx -0.999$ and $\mathcal{C}_D(\mathbf{r}_2) \approx +0.999$, respectively. }
\label{figure:junction}
\end{figure}

\subsection{The numerical results}\label{section:numerics}

The real-space Chern number is represented in Fig. \ref{fig:local}, for an approximant of order $\mathcal{O}=12$, with a magnetic flux per tile $\phi/\phi_0=1/3$, and a Fermi energy $E_{\text{F}}=-1.5 J$ located within the first bulk gap. We observe that the local quantity $\mathcal{C} (\mathbf{r}_j) \approx -1$ only slightly fluctuates within the bulk of the iGRT, which we attribute to the  regularity of the quasicrystal, see Fig. \ref{fig:local}. In order to inhibit any local fluctuation effects, which could potentially arise in more irregular quasicrystals, we measure the topological order in the bulk by averaging the real-space Chern number over a disk $D$ of radius $r_D$, which is chosen such as to cover the bulk only. The resulting quantity, denoted $\mathcal{C}_D$, is found to be equal to $\mathcal{C}_D \approx -1.002$ for the situation shown in Fig. \ref{fig:local}. Close to the edge, the real-space Chern number $\mathcal{C} (\mathbf{r}_j)$ strongly fluctuates, and reaches arbitrary high values that depend on the system size. Thus, this quantity no longer characterizes the topological property of the system in this boundary region.

We further investigate the local property of the real-space Chern number $\mathcal{C} (\mathbf{r}_j)$, by considering a heterojunction within the Rauzy tilling of order $\mathcal{O}=13$. Each region of the junction is subjected to a different magnetic flux, as illustrated in Fig. \ref{figure:junction} (a). The fluxes are chosen to be $\phi=1/3$ and $\phi=2/3 \equiv - 1/3$, respectively, so that the two regions are associated with opposite Chern numbers. Setting the Fermi energy within the first bulk gap, $E_{\text{F}}=-1.5J$, each region corresponds to a Chern insulating phase with Chern number $\nu_{\text{tot}}=-1$ and $\nu_{\text{tot}}=+1$, respectively. The result depicted in Fig. \ref{figure:junction} (b) shows that these topological invariants are well captured by the (local) real-space Chern number $\mathcal{C} (\mathbf{r}_j)$, which indeed distinguishes between the two topologically-ordered regions. The averaged Chern number $\mathcal{C}_D(\mathbf{r}_0)$, defined in Eq. \eqref{lattice_smooth}, has been evaluated within the two regions, considering two disks $D$ of radius $r_D \approx 4 a$ (i.e.~containing 98 sites each), yielding $\mathcal{C}_D(\mathbf{r}_1) \approx -0.999$ and $\mathcal{C}_D(\mathbf{r}_1) \approx +0.999$, respectively. The fluctuations of the real-space Chern number $\mathcal{C} (\mathbf{r}_j)$ are found to be of the order of $10^{-2}$ within the bulk of these regions. Note that such a result could not be easily obtained through momentum-space-integration methods \cite{Fukui:2005ua}, which would require to set periodic boundary conditions unto the two regions, separately, to evaluate the corresponding Chern numbers $\nu_{\text{tot}}=\pm1$.

\begin{figure}[!]
\begin{center}
\includegraphics[width=8cm]{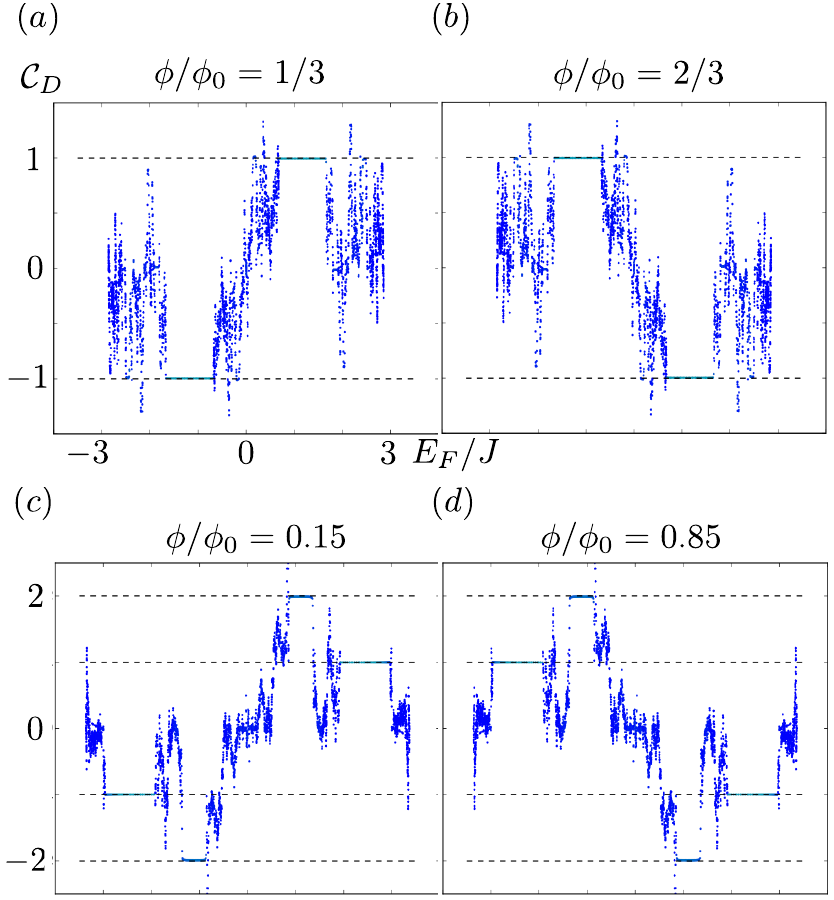},
\end{center}
\caption{Plot of the averaged Chern number $\mathcal{C}_D$ as a function of the Fermi energy $E_{\text{F}}$  for an approximant of order $\mathcal{O}=13$ (3136 sites), and for four values of the magnetic flux per tile, see (a)-(d). Each plotted point is colored (blue to green) according to the value of the edge-locality marker $\mathfrak{B}_{\lambda}$ for $E_{\text{F}} = E_{\lambda}$. The averaged Chern number $\mathcal{C}_D$, obtained by averaging the real-space Chern number in a disk $D$ containing 187 sites, is found to be close to the topological Chern numbers $\nu_{\text{tot}}=\pm 1, \pm 2$, when $E_{\text{F}}$ falls into the different low-DOS spectral ranges.}
\label{fig:cond}
\end{figure}

The topology associated with the different gaps of the butterfly-like spectrum in Fig.~\ref{fig:butterfly_density} can be estimated by computing the averaged real-space Chern number $\mathcal{C}_D$ for various values of the Fermi energy $E_{\text{F}}$ and magnetic flux $\phi$. The result is shown in Fig.~\ref{fig:cond}, where $\mathcal{C}_D$ is plotted as a function of the Fermi energy for various values of the flux $\phi$.   We choose the radius $r_D \approx 6 a$ (containing 187 sites) for the approximant $\mathcal{O}=13$, which already provides a good evaluation of the topological Chern number $\nu_{\text{tot}}$ at the center of the quasicrystal. As already stated above, we find that the regular pattern of the iGRT (i.e. the unique tile area $A_\text{tile}$ and the uniform magnetic flux $\phi$ over the entire quasicrystal) leads to weak fluctuations of the real-space Chern number $\mathcal{C} (\mathbf{r}_j)$ in Eq. \eqref{lattice_chern}. By setting the flux to the values $\phi/\phi_0=1/3$ and $\phi/\phi_0=2/3$, one obtains that the two main gaps of the butterfly are indeed characterized by the topological index $\mathcal{C}_D \approx \nu_{\text{tot}} =\pm 1$, respectively [see Fig. \ref{fig:butterfly_density} (a)-(b)].  Smaller gaps are probed by setting the flux to the values $\phi/\phi_0=0.15$ and $\phi/\phi_0=0.85$, where new gaps associated with the total Chern numbers $\nu_{\text{tot}} =\pm 2$ are clearly probed.  This analysis fully confirms the presence of various Chern insulating phases in the quasicrystal, whose topological phase diagram follows the shape of the butterfly spectrum presented in Fig. \ref{fig:butterfly_density}.


\section{Conclusions}\label{section:conclusion}

This work aimed to demonstrate the existence of Chern insulating phases in 2D quasicrystals, based on a study of the iGRT subjected to a uniform magnetic field. Such transport properties of non-periodic lattice systems could be investigated in real solid-state materials, for instance, generalizing the construction and analysis of moiré superlattices \cite{Dean:2013bv}. Quasicrystal potentials could also be designed for cold atoms moving in optical lattices, for instance, using the novel technology of light-intensity masks \cite{Gaunt:2013PRL,Corman:2014cm,Blochprivate}. In this configuration, the magnetic fields should be produced externally, e.g. through atom-light coupling \cite{Dalibard2011,Goldman:2014RPP}, or more simply, through time-modulation protocols \cite{Jotzu:2015kz,Aidelsburger:2014,Struck2013NatPhys,Goldman:2014PRX,Bukov:2014,Goldman:2014arxiv,Arimondo:2012AP}. Besides, photonic crystals also constitute a versatile platform to investigate the topological properties of exotic lattice structures \cite{Hafezi:2011NatPhys,Verbin:2013,Rechtsman:2013fe,Hafezi:2013NatPhot}. The present work opens an interesting route towards the observation of topological properties in a wide range of (non-periodic) lattice structures. In this regard, we notice that, beyond the iGRT, there exist other quasicrystals, which exhibit a gapped energy spectrum even in the absence of a magnetic field \cite{Sire:1989,SireMosseri:1990,SireBellissard:1990,BenzaSire:1990}. Of particular interest would be the possibility to reveal topological insulators in quasicrystals exhibiting large spin-orbit coupling, both in 2D and 3D materials.  

\begin{acknowledgments}

The authors are pleased to acknowledge I. Bloch, J. Dalibard, M. Lewenstein, M. A. Martin Delgado, P. Massignan, M. M\"uller, S. Nascimb\`ene, and L. Tarruell  for valuable discussions. DTT, AD and NG are supported by the FRS-FNRS (Belgium). This research is also financially supported by the Belgian Federal Government under the Interuniversity Attraction Pole project P7/18 ``DYGEST". \\
\end{acknowledgments}



%





\end{document}